\documentclass[aps,prl,superscriptaddress,amsmath,amssymb,floatfix,twocolumn]{revtex4}

\usepackage{graphicx}
\usepackage{epstopdf}
\usepackage{dcolumn}
\usepackage{bm}
\usepackage{color}
\usepackage{ulem}
\usepackage{epsfig,float,afterpage,amssymb,wrapfig,psfrag}
\usepackage[caption=false]{subfig}
\usepackage{placeins}
\usepackage{comment}
\usepackage{float}

\DeclareMathAlphabet{\bi}{OML}{cmm}{b}{it}

\newcommand{\be}[0]{\begin{equation}}
\newcommand{\ee}[0]{\end{equation}}
\newcommand{\ba}[0]{\begin{eqnarray}}
\newcommand{\ea}[0]{\end{eqnarray}}

\begin{document}
\title{Quantum Criticality and Dynamical Kondo Effect in an SU(2) Anderson Lattice Model}

\author{Haoyu Hu}
\email{hh25@rice.edu}
\affiliation{Department of Physics and Astronomy, Rice Center for Quantum Materials,  Rice University,
Houston, Texas, 77005, USA}

\author{Ang Cai}
\affiliation{Department of Physics and Astronomy, Rice Center for Quantum Materials,  Rice University,
Houston, Texas, 77005, USA}

\author{Qimiao Si}
\email{qmsi@rice.edu}
\affiliation{Department of Physics and Astronomy, Rice Center for Quantum Materials,  Rice University,
Houston, Texas, 77005, USA}

\begin{abstract}
Metallic quantum criticality often develops in strongly correlated systems 
with local effective degrees of freedom.
In this work, we
consider an Anderson lattice model with $SU(2)$ symmetry. 
The model is treated by the extended dynamical mean-field theory (EDMFT) in combination with a 
continuous-time quantum Monte Carlo method.
We demonstrate  a continuous quantum phase transition,
establish the ensuing quantum critical point to be of a Kondo-destruction type,
and determine the anomalous scaling properties.
We connect the continuous nature of the transition to a dynamical Kondo effect, 
which we characterize in terms of a local entanglement entropy and related properties.
This effect elucidates the unusual behavior of quantum critical heavy fermion systems.
\end{abstract}

\date{\today}

\maketitle

{\it Introduction. }
The nature of quantum critical points (QCPs), especially in metallic systems,
is of extensive interest  to a variety of strongly correlated
systems\cite{JLTP2010,Kirchner_rmp20,KeimerMoore,Coleman-Nature,Sachdev-book}.
Strong correlations often produce local effective degrees of freedom as a part of the building blocks for the low-energy physics. This is exemplified by antiferromagnetic (AFM) heavy fermion metals,
in which local moments couple and interplay with itinerant electrons. The fate of the local moments
and associated
Kondo effect has played a central role in elucidating the AFM heavy fermion QCPs,
both theoretically\cite{Si-Nature,Colemanetal} and 
experimentally\cite{Prochaska20,schroder,paschen2004,shishido2005,Par06.1}. 
Of particular importance is the notion of Kondo destruction\cite{Kirchner_rmp20}, 
which 
corresponds to the disintegration of heavy quasiparticles.
A Kondo-destruction QCP amounts to a delocalization-localization
transition of the underlying $f$-electrons, 
thereby involving a sudden reconstruction of the Fermi surface.
It also makes the quasiparticle weight at the QCP to vanish on the entire Fermi surface, 
which is responsible for the strange metal behavior in the quantum critical regime 
and the divergence of the effective carrier mass.
Similar features may develop at the Mott transition and doped Mott 
insulators\cite{Senthil08.1,Ter11},
in light of the indications for a divergent carrier mass and a Fermi surface reconstruction in the 
high $T_c$ cuprates 
near their optimal hole doping\cite{Ram15.1,Bad16.1}.
 
The AFM quantum phase transitions in heavy fermion metals result from a competition between
the Kondo and Ruderman-Kittel-Kasuya-Yosida (RKKY) couplings.
These interactions respectively promote a Kondo-screened paramagnetic ground state
and a long-range AFM order.
Kondo destruction was initially studied by analyzing the fate of the Kondo effect near AFM QCPs
\cite{Si-Nature,local_fluct_qm_2003}.
In the case of Ising-anisotropic Kondo lattice models, the continuous nature of the quantum phase transition
has been demonstrated in a number of studies\cite{si2014kondo,jxzhu2003,glossop2007prl,jxzhu2007}. 
While some of the quantum critical heavy fermion
systems are Ising-anisotropic\cite{JLTP2010,schroder}, others have a continuous spin 
symmetry\cite{paschen2004,shishido2005,Par06.1,Das_prl14}. It is thus important to
address the issue in Kondo or Anderson
lattice models with continuous spin symmetry. The latter is also important when
connections are explored between the quantum criticality of the Kondo systems with that of the Mott-Hubbard systems such as the cuprates, which are to a good approximation $SU(2)$ symmetric.

In this Letter, we consider an Anderson lattice model with an $SU(2)$ symmetry.
We study the lattice model in terms of a self-consistent 
Bose-Fermi Anderson model via the
extended dynamical mean-field theory (EDMFT)\cite{si1996kosterlitz,Smith_edmft,chitra2000effect}.
Our study has become possible due to the recent development of a continuous-time 
Quantum Monte Carlo (CTQMC) approach
suitable for the $SU(2)$-symmetric Bose-Fermi Anderson model\cite{Ang_su2_bf}, which built on
 the general CTQMC 
method\cite{Gull11,Otsuki_su2_bf}; importantly, we reach temperatures lower than
$10^{-3}$ of the bare Kondo temperature.
We show that the quantum phase transition is continuous. For the ensuing QCP,
we  demonstrate its Kondo-destruction nature,
and find that the spin dynamics obey $\omega/T$ scaling.
 Equally important, we establish a dynamical Kondo effect 
 by calculating the local entanglement entropy\cite{EE_KD, EE_KD_long_range}
 as well as the cross correlations between the local moment and conduction-electron spins. 
 The dynamical Kondo effect
underlies the continuous nature of the quantum phase transition,
and elucidates the unusual properties of quantum critical heavy fermion systems.

{\it Model and method. }
We study the Hamiltonian
\begin{eqnarray}
H&=& (U/2) \sum_i \bigg[\sum_{\sigma} d_{i,\sigma}^\dag d_{i,\sigma} -1 \bigg]^2 + 
\sum_{ij} I_{ij}\bm{S}_i \cdot \bm{S}_j 
\nonumber \\ 
&&+V\sum_{i,\sigma}[ c_{i,\sigma}^\dag d_{i,\sigma} +d_{i,\sigma}^\dag c_{i,\sigma}]
+\sum_{p,\sigma} \epsilon_p c_{p,\sigma}^\dag c_{p,\sigma}  \,.
\label{eq:ham}
\end{eqnarray}
Here, $c_{i,\sigma}^\dag$ ($d_{i,\sigma}^\dag$) creates 
a conduction $c$ (local $d$) electron of spin $\sigma$ at site $i$,
and  $\bm{S}_{i} = (d_{i}^\dag \bm{\sigma} d_i)/2$ (with $\bm{\sigma}$ being the Pauli matrices)
 represents the spin 
  of 
the $d$-electrons (denoting the physical $f$-electrons).
The hybridization $V$ couples
the $c$-electrons, which has a dispersion $\epsilon_p$,
and the $d$-electron spins,
which involve an AFM RKKY interaction $I_{ij}$.
Finally, the repulsive Hubbard interaction
$U$ is responsible for 
turning 
the $d$-electrons into local moments;
when $U$ is sufficiently large, the model is equivalent to a Kondo lattice Hamiltonian, with an effective 
AFM Kondo coupling $J_K$ being 
second order in $V$.

The EDMFT approach\cite{si2014kondo,si1996kosterlitz,Smith_edmft,chitra2000effect}
takes into account the {\it dynamical}
competition between the hybridization/Kondo and RKKY interactions.
Here, the lattice Hamiltonian Eq.~\ref{eq:ham} 
is solved in terms of a self-consistent 
Bose-Fermi Anderson model.
In the latter,
the local $d$ electrons couple to a fermionic bath and a bosonic bath, 
where the fermionic bath comes from the conduction electrons and the bosonic bath represents the fluctuations of local moments\cite{supp}.
After integrating out both baths, we reach the following action:
\begin{eqnarray}
&&S_{BFA}\nonumber\\
&=&  \int_0^\beta d\tau\bigg[ \sum_{\sigma} d_{\sigma}^\dag \partial_\tau d_{\sigma} +\frac{U}{2}( n_{d,\uparrow}+n_{d,\downarrow}-1)^2 +h_{loc}\ S^z \bigg]  \nonumber \\
&&-\int_0^\beta d\tau 
 d\tau'
 \, [ \,  \sum_{\sigma} d_{\sigma}^\dag(\tau)V^2G_{c}(\tau-\tau') d_{\sigma}(\tau')  \nonumber \\
&&
+ (1/2) \sum_{\alpha\in\{x,y,z\}}S^{\alpha}(\tau)[\chi_{0}^{\alpha}]^{-1} (\tau-\tau')S^{\alpha}(\tau')
\, ]
\, ,
\label{eq:imp}
\end{eqnarray}
where $\beta=1/T$,
$h_{loc}$ is a static Weiss field, which captures the 
 AFM order, while
$G_c$ and $\chi_0$ denote the Green's functions of the fermionic and bosonic bath\cite{supp}, respectively. 
The self-consistency conditions are:
\begin{eqnarray}
&&\chi_{loc}^\alpha(i\omega_n) = \int_{-\infty}^{\infty} d\epsilon 
\rho_I(\epsilon)/[\epsilon+M^\alpha(i\omega_n)]
\nonumber \\
&&~~~~
 M^\alpha(i\omega_n) = [\chi_{0}^{\alpha}]^{-1} (i\omega_n) +  1/{\chi^\alpha_{loc}(i\omega_n)}  \nonumber \\
&&G_{c}(i\omega_n) = \int_{-\infty}^{\infty} d\epsilon 
\rho_0(\epsilon)/[-i\omega_n + \epsilon +\Sigma_c(i\omega_n) ] \nonumber \\
&&h_{loc} = -[2I- [\chi_{0}^{\alpha}]^{-1} (i\omega_n=0)
]m_{AF} 
 \,.
\label{eq:self_cons}
\end{eqnarray}
Here, $\alpha \in \{x,y,z\} $ represents the spin components, with the magnetic order
taken along 
$\alpha=z$;
$\rho_I(\epsilon)$ denotes the RKKY density of states,
which is
obtained from $\rho_I(\epsilon) = \sum_{\bm q} \delta(\epsilon-I_{\bm q})$.
The RKKY interaction 
$I_{\bm q}$ 
is
the Fourier transformation of $I_{ij}$,
and 
is the most negative at the AFM wave vector ${\bm Q}$;
we consider
 $I_{\bm Q}=-2I$ and the
 density of states
 $\rho_I(\epsilon) = \theta(2I-|\epsilon|)/(4I)$, 
 which incorporates two-dimensional 
magnetic fluctuations\cite{Si-Nature}.
 The  irreducible (and momentum-independent) quantity
$M^\alpha(i\omega_n)$ 
reflects a spin self-energy\cite{si1996kosterlitz,Smith_edmft}.
The  local irreducible spin susceptibility $\chi_{loc}$ and 
ordered moment $m_{AF}$ are:
\begin{eqnarray}
\chi_{loc}^\alpha(\tau) &=&\langle  T_{\tau} : S^\alpha:(\tau) :S^\alpha:(0) \rangle_{S_{BFA}} \nonumber \\
m_{AF} &=& \langle  S^z \rangle_{S_{BFA}}  \, .
\label{eq:eve}
\end{eqnarray}
The expectation value $\langle \cdot \cdot \cdot \rangle_{S_{BFA}}$ 
is taken with respect to the action $S_{BFA}$ (Eq.~\ref{eq:imp}),
and the normal-ordered operators are 
$:S^z: \equiv S^z -\langle S^z \rangle_{S_{BFA}}$ and
$:S^{x,y}: \equiv S^{x,y}$.
We consider 
a generic electron filling\cite{Si_2005}
with
 the conduction-electron band
 having a nonzero density of states at the zero energy
in terms of a featureless $\rho_0(\epsilon)$, for which
$G_c(i\omega_n) = \frac{1}{2D} \log(\frac{-i\omega_n +D}{-i\omega_n -D})$.
The equations are iterated until
convergence, which corresponds 
 to 
the differences between the two iterations being less than 0.1\%.

{\it Continuous AFM quantum phase transition.}
\begin{figure}[t!]
\captionsetup[subfigure]{labelformat=empty}
  \centering
    \mbox{\includegraphics[width=0.98\columnwidth]{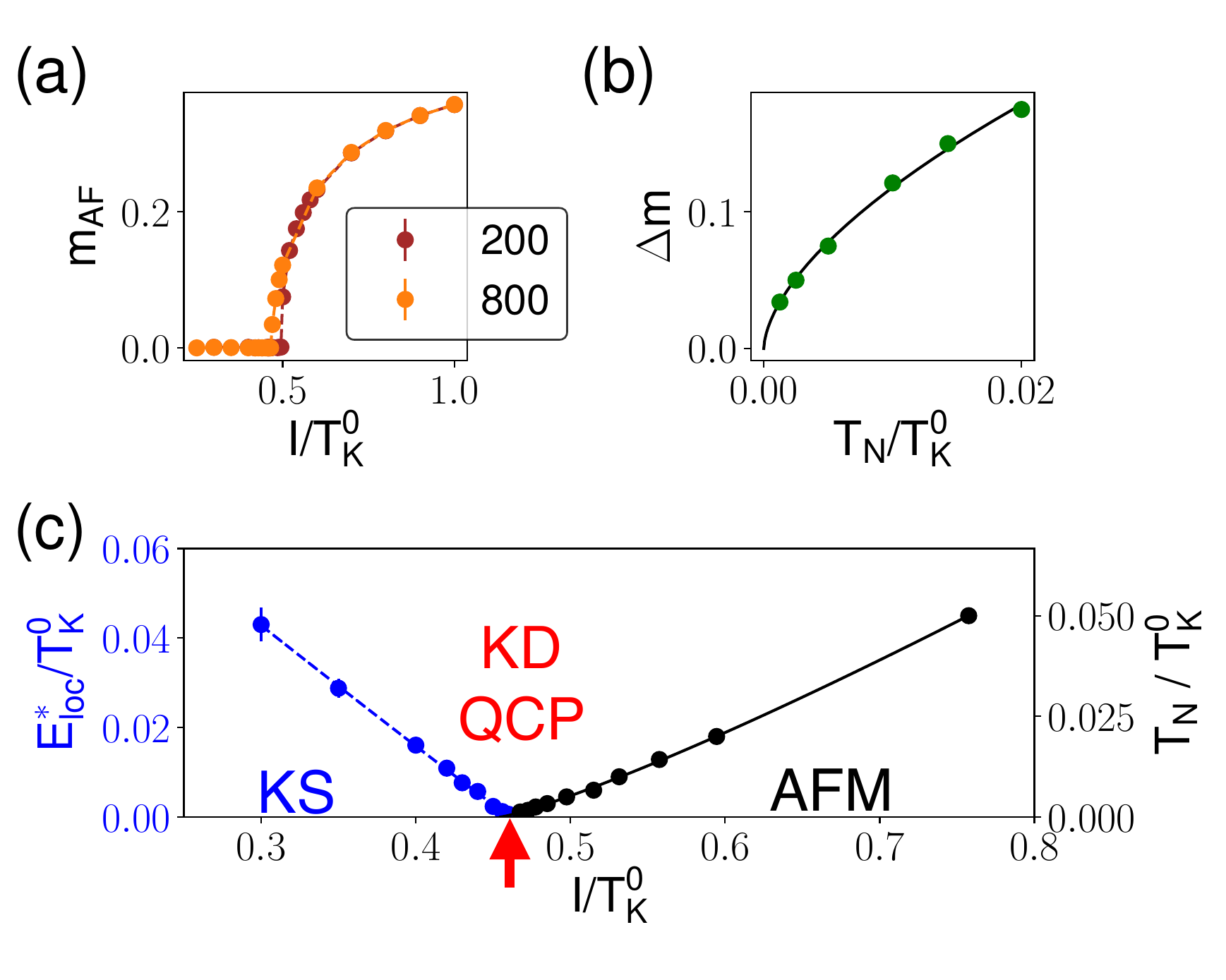}}    
\caption{(Color online) (a) Evolution of the AFM order parameter 
$m_{AF}$ with the ratio
$I/T_K^0$, at inverse temperatures 
$\beta T_K^0=200,800$.
(b) Jump in the order parameter $\Delta m$ {\it vs.} the N\'eel temperature $T_{N}$,
 extrapolating to zero
 as $T_{N} \rightarrow 0^+$.
(c) Phase diagram, showing also the Kondo-destruction (KD) energy scale $E_{loc}^*$ {\it vs.}
$I/T_K^0$.
 KS and AFM respectively denote Kondo-screened and antiferromagnetic phases.
 The red arrow marks the QCP at $I=I_c$. }
\label{fig:phase_diagram}
\end{figure}
To determine the phase diagram, we work with the generic parameters
$U = 0.25, V=0.40, D=1.0$, for which the
bare Kondo temperature $T_K^0=1 $. 
Hereinafter, we only vary the RKKY interaction $I$ to tune the ratio 
of the RKKY interaction to the bare Kondo scale,
$I/T_K^0$. 

We perform calculations at various
values of
 temperature $T$ and the
 tuning parameter
$I/T_K^0$. 
For a given $T$, an isothermal AFM phase transition
is seen through the onset of
the order parameter
$m_{AF}$, as illustrated in Fig.\,\ref{fig:phase_diagram}(a).
We observe a jump in the order parameter, $\Delta m$,
at the isothermal transition point, indicating that the finite-temperature phase transition
is first order\cite{jxzhu2003}.
The phase diagram is shown 
in Fig.\,\ref{fig:phase_diagram}(c). 
As we approach the zero-temperature phase transition 
along the  
$T_N$
line (black solid curve), 
 $\Delta m$ 
 decreases 
 as shown in Fig.\,\ref{fig:phase_diagram}(b);
  within the error bar, it extrapolates to zero in the zero temperature limit.
 This establishes 
a continuous 
quantum phase transition at $I=I_c$.

{\it Kondo destruction, quantum criticality and anomalous scaling. }
\begin{figure}[t!]
\captionsetup[subfigure]{labelformat=empty}
  \centering
    \mbox{\includegraphics[width=0.90\columnwidth]{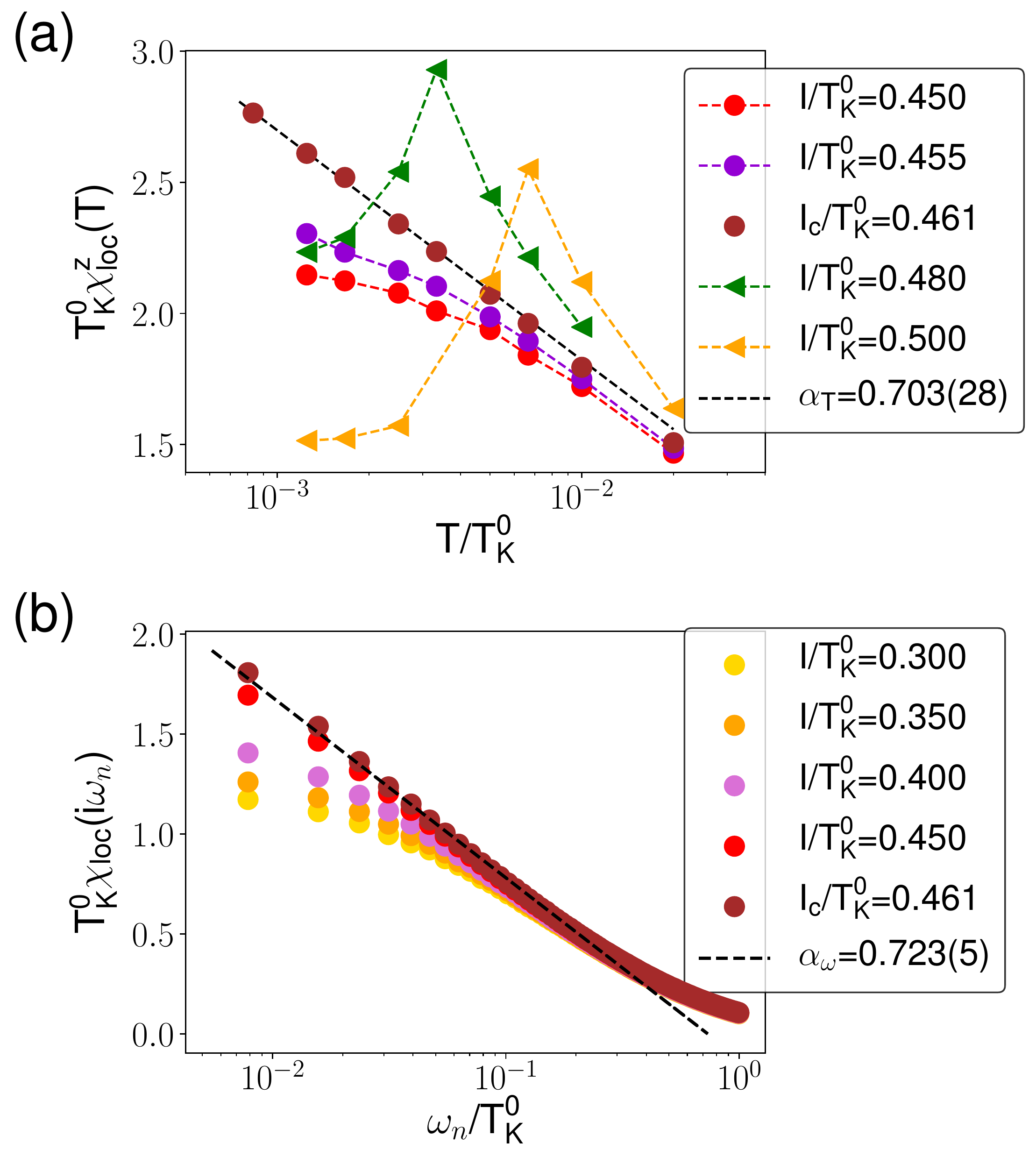}}    
\caption{(Color online) (a) Temperature dependence of the local spin susceptibility $T_K^0\chi^z_{loc}(T)$.
The QCP is at $I_c/T_K^0=0.461$ (see Fig.\,S2\cite{supp}).
(b) The dynamical local spin susceptibility in the paramagnetic phase $(I\le I_c)$
at inverse temperature $\beta T_K^0=800$. }
\label{fig:local_spin_sus}
\end{figure}
We now turn to the  nature and properties of the QCP.
We focus on the $\alpha=z$ component of the spin susceptibility, 
which captures the AFM order and, outside of the ordered region,
is equal to the 
other two components: we will use $\chi^z$ to denote this quantity inside of the AFM order and simply $\chi$ outside of it.

The temperature dependence of the static local spin susceptibility 
$\chi_{loc}(T)$, defined as $\chi_{loc}(i\omega_n=0,T)$, 
is presented in Fig.\,\ref{fig:local_spin_sus}(a) (as well as Fig.\,S1\cite{supp}) for various values of 
the RKKY interaction $I$. 
For $I>I_c$, it shows a peak at $T_N$. For $I<I_c$, it saturates to a finite value at low temperatures, signifying Kondo screening.
$\chi_{loc}(T\rightarrow 0)$ is divergent at $I=I_c$, the QCP. This means that, at the AFM QCP,
the Kondo effect is placed at the critical Kondo-destruction point,
as seen from the renormalization-group (RG) flow of the Bose-Fermi Kondo 
model\cite{si1996kosterlitz,smith1999,sengupta2000,zhu2002,zarand2002}
(summarized in supplementary materials\cite{supp}, Fig.\,S2; particularly the dashed arrow).
More precisely, $\chi_{loc}(T)$ at the QCP is logarithmically singular, 
corresponding to the bosonic bath of the Bose-Fermi Anderson
model having a sub-ohmic spectrum with its power-law exponent $0^+$ (Ref.\,\cite{Si-Nature}).
We can then express $\chi_{loc}(T)$ in the following form \cite{local_fluct_qm_2003}:
\be 
\chi_{loc}(T)=-\frac{\alpha_T}{4I}\log(T)+b_T  \, .
\label{eq:log_T}
 \ee 

In Fig.\,\ref{fig:local_spin_sus}(b), we show the dynamical local spin susceptibility $\chi_{loc}(i\omega_n)$ 
in the paramagnetic part of the phase diagram.
At the critical point, $I=I_c$,
it also is found to be singular and satisfy
\be
\chi_{loc}(i\omega_n) = -\frac{\alpha_\omega}{4I} \log(\omega_n) +b_\omega \,
\label{eq:log_w}
\ee
in a large dynamical range ( $0.004 T_K^0 \le\omega_n \lessapprox 0.300 T_K^0  $). 
Away from the critical point,  $I>I_c$,
the local susceptibility saturates, which places the local Kondo problem to be on the Kondo-screened side 
of the RG flow (supplementary materials\cite{supp}, Fig.\,S2).
We can introduce a local Kondo energy scale $E_{loc}^*$ to characterize this saturation\cite{Gre03.2,jxzhu2003},
by fitting 
 $\chi_{loc}(i\omega_n)$ in terms of $ A + B \log( \omega_n+ E_{loc}^* )$.
 The resulting $E_{loc}^*$ as a function of the tuning parameter $I/T_K^0$ is already shown in 
 Fig.\,\ref{fig:phase_diagram}(c). This energy scale collapses as the system approaches the QCP, 
 manifesting  the
 Kondo destruction at the QCP. 

\begin{figure}[t!]
\captionsetup[subfigure]{labelformat=empty}
  \centering
    \mbox{\includegraphics[width=0.90\columnwidth]{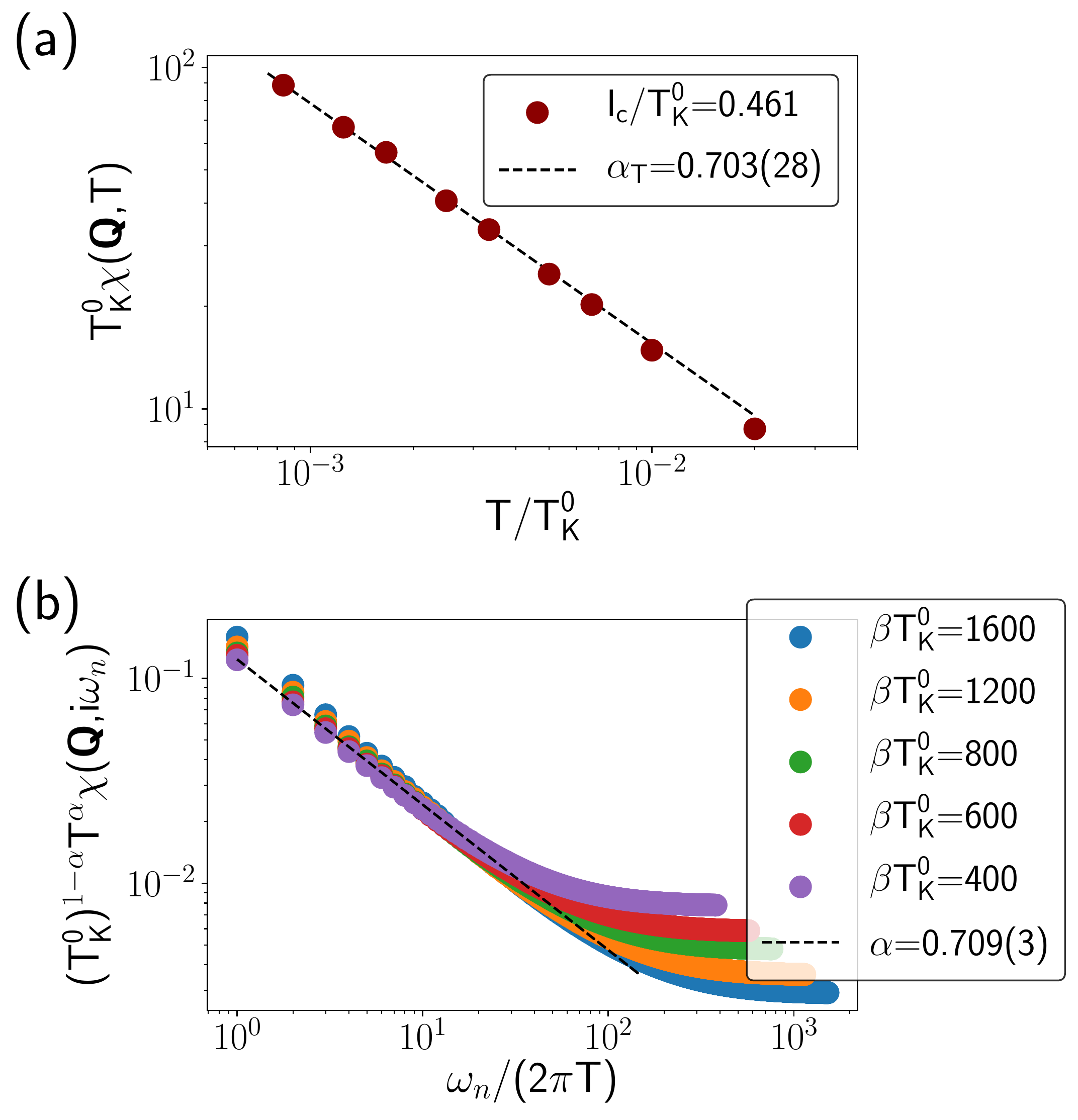}}    
\caption{(Color online) (a) Static lattice spin susceptibility vs. temperature at $I=I_c$.
 (b) Demonstration of $\omega_n/T$ scaling 
 for the dynamical lattice spin susceptibility at $I=I_c$. }
\label{fig:w_T_scaling}
\end{figure}

These results also imply that, at the QCP, $I=I_c$,  the AFM spin susceptibility has the following power-law
temperature and frequency dependences\cite{supp}:
\ba 
&&\chi({\bf Q},T) \propto T^{-\alpha_T} \, , \nonumber \\
&&\chi({\bf Q},i\omega_n) \propto \omega_n^{-\alpha_\omega} \,.
\label{eq:critical_T}
\ea
The static lattice susceptibility $\chi({\bf Q},T)$ is shown in Fig.\,\ref{fig:w_T_scaling}(a),
and its dynamical counterpart, at a low-temperature $T=1.125 \times 10^{-3} T_K^0$,
is presented in Fig.\,S3.
We find the critical exponents 
to be {$\alpha_T=0.701(28)$ and $\alpha_\omega=0.723(5)$}.

The temperature and frequency exponents have the same fractional value 
within the numerical uncertainly.
This result suggests 
that the AFM dynamical spin susceptibility obeys $\omega/T$ scaling in the quantum critical regime.
To address this issue further, we calculate the AFM dynamical spin susceptibility at the QCP, $I=I_c$,
as a function of both
 frequency and temperature over a range of low temperatures ($T_K^0/1600 < T < T_K^0/400$).
In Fig.\,\ref{fig:w_T_scaling}(b), 
we show that $(T_K^0)^{1-\alpha} T^{{\alpha}}\chi (Q,\omega_n,T)$ 
at the various temperatures collapses in the form of $\omega_n/T$ scaling.
The critical exponent $\widetilde{\alpha}$ is unbiasedly determined by
this procedure. Its value, $\alpha=0.709(3)$, is compatible with $\alpha_T$ and ${\alpha_\omega}$ described earlier.
Together, these results demonstrate that the dynamical spin susceptibility displays $\omega/T$  scaling and fractional scaling 
exponents.

The usual spin-density-wave
QCP\cite{hertz1976quantum,millis1993effect,moriya2012spin} 
falls within the Landau framework of order-parameter fluctuations and 
corresponds to a Gaussian fixed point.
The $\omega/T$ scaling we found signifies an interacting fixed point. The collapse of the 
energy scale $E_{loc}^*$ at the QCP signifies that the Kondo destruction underlies the beyond-Landau physics.
In turn, this implies that 
the quasiparticle weight vanishes as the QCP is approached, and the Fermi surface jumps between large 
(counting the Kondo resonance) and 
small (not counting the Kondo resonance) across the QCP.

\begin{figure}[t!]
\captionsetup[subfigure]{labelformat=empty}
  \centering
    \mbox{\includegraphics[width=0.98\columnwidth]{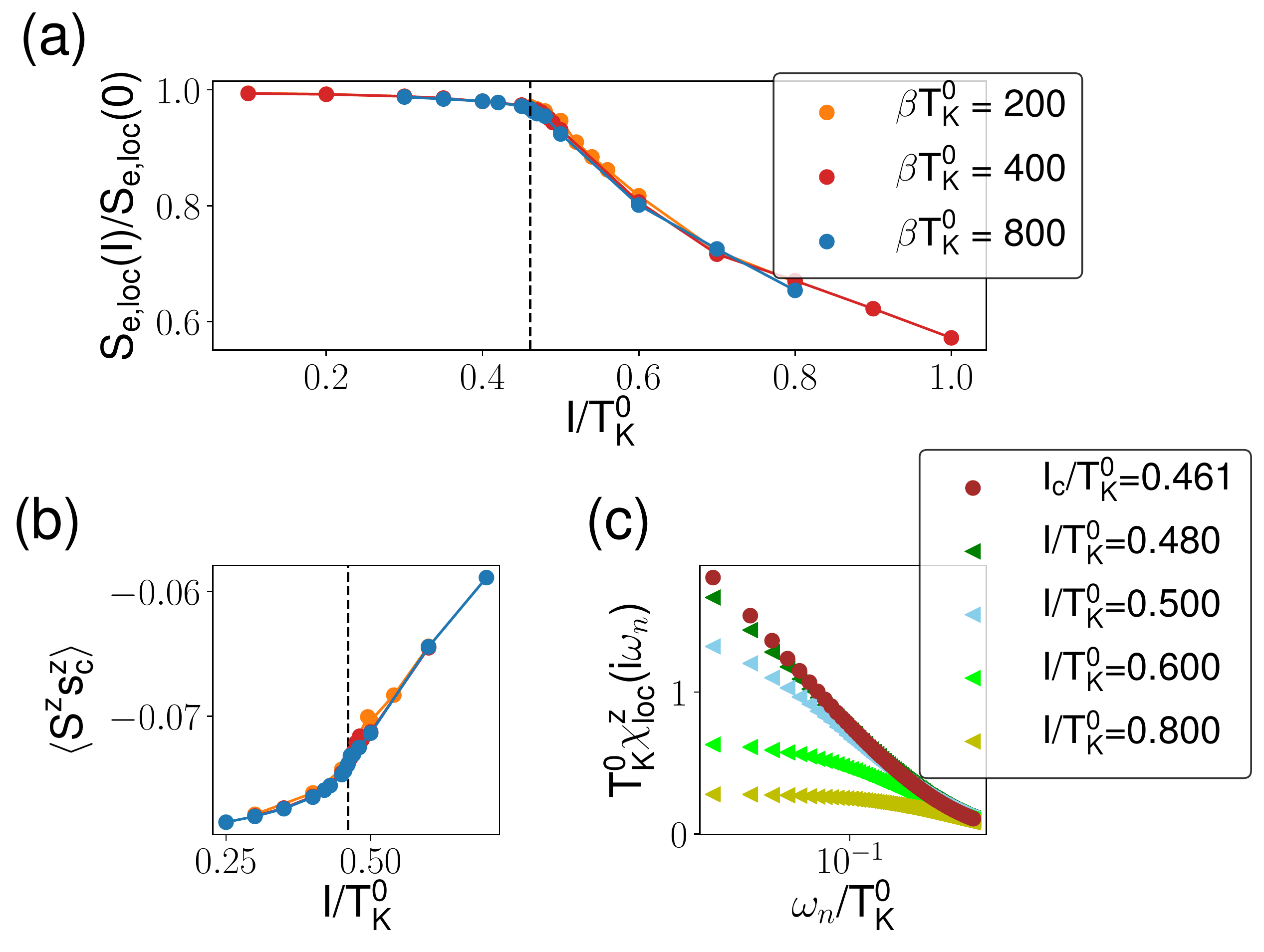}}    
\caption{(Color online) (a) Evolution of the local entanglement entropy with the tuning parameter $I/T_K^0$.
The vertical dashed line marks the AFM QCP.
(b) $\langle S^z s_c^z \rangle$ vs. the tuning parameter across the QCP.
(c) Dynamical local susceptibility for $I\ge I_c$ and $\beta T_K^0 =800$.}
\label{fig:dynamic_kondo}
\end{figure}

{\it Dynamical Kondo effect. } 
We now turn to understand why the quantum phase transition is continuous 
when the heavy quasiparticles disintegrate at the transition.
To do so, we first calculate the entanglement entropy of a local $d$ electron.
The entanglement property has been calculated in the standing-alone Bose-Fermi Kondo impurity
models\cite{EE_KD, EE_KD_long_range},
but has not been studied in any Kondo/Anderson lattice models. The local entanglement entropy is defined as 
$S_{e,loc} = -\text{Tr}[\rho_{loc} \log (\rho_{loc})]$,
where $\rho_{loc}$ is the density matrix of the 
lcoal $d$ electron\cite{EE_loc}. 
In Fig.\,\ref{fig:dynamic_kondo}(a), we show the evolution of $S_{e,loc}$ 
across the QCP.
In the Kondo-screened phase, due to the Kondo effect,
 the local $d$ electrons and the conduction-electron band 
are Kondo entangled, which results in a large $S_{e,loc}$  [we find $S_{e,loc}(I=0)= 1.386$].
Increasing the RKKY interaction $I$ through the QCP, $I_c$,
$S_{e,loc}$ drops precipitously, capturing the Kondo destruction.
Importantly, it stays nonzero inside the Kondo-destroyed phase at $I>I_c$. This implies that residual Kondo-singlet correlations persist. 
This is to be contrasted with the naive mean-field picture,  which would have associated 
the Kondo destruction with a complete decoupling between the local $d$ electron and conduction-electron bands;
in that picture, the entanglement entropy must vanish.
We interpret our result as signifying the persistence of the dynamical Kondo-singlet correlations 
in the Kondo-destroyed phase, even though the static Kondo-singlet amplitude has vanished in the ground state.

This point can be further demonstrated by the cross correlation
between the local moment and conduction-electron spins, which 
has recently been considered in
quantum impurity models\cite{dynamical_kondo}.
The expectation value of $\langle S^z s_{c}^z \rangle $ 
is shown in Fig.\,\ref{fig:dynamic_kondo} (b). While it is naturally nonzero in the Kondo-screened phase
at $I<I_c$, it remains so at the Kondo-destruction QCP ($I=I_c$) and in the Kondo-destroyed phase ($I>I_c$).
Through Kramers-Kronig relation\cite{dynamical_kondo}, our result, derived for the first time in any lattice model,
implies the persistence of the dynamical Kondo correlations across the QCP and into the Kondo-destroyed phase.

The dynamical Kondo effect can be further illustrated by the irreducible local spin susceptibility,
$\chi_{loc}(i\omega_n)$,
in the ordered phase. 
The result is given in Fig.\,\ref{fig:dynamic_kondo} (c).
If the local $d$ electrons were completely decoupled from the conduction-electron band,
$\chi_{loc}(i\omega_n)$ would vanish at nonzero frequencies. Indeed,
 $\chi_{loc}(i\omega_n)$ is negligible at $I \gg I_c$, {\it i.e.} deep into the ordered phase.
As $I$ is reduced towards $I_c$, $\chi_{loc}(i\omega_n)$ progressively grows, 
reflecting the increased dynamical Kondo-singlet correlations
in the Kondo-destroyed phase.

The vanishing of the spectral weight for the well-defined Kondo resonance
as the transition is approached from the paramagnetic side underlies the sudden jump of the Fermi surface across the QCP. In a naive mean-field picture, one would have expected the transition to be first order.
The dynamical Kondo effect we have demonstrated 
underlies the continuity of the Kondo correlations across the 
transition and into the Kondo-destroyed phase, which makes possible for the quantum
phase transition to be continuous.

{\it Discussion.}
Several remarks are in order.
First, the dynamical Kondo effect captures the quantum fluctuations in the Kondo-destroyed 
phase. One of the unusual properties observed\cite{Mar18.1,Geg02.1}
in the prototype heavy fermion metals with 
Kondo-destruction QCPs is the pronounced quasiparticle mass in the Kondo-destroyed phase. The quantum fluctuations associated with the 
dynamical Kondo effect provide a natural understanding of this property.

Second, our EDMFT-based results for the beyond-Landau QCP in the $SU(2)$ Anderson lattice model set the stage
to connect with what happens in the $SU(2)$ Hubbard-Heisenberg models. Recently, EDMFT analyses have 
also been carried out in the latter models\cite{Joshi20,Cha20,Tarnopolsky20}, and have implicated related
QCPs. Since the effective local problem in that case is quite similar to what we analyze for the $SU(2)$ Anderson 
lattice model, it will be instructive to see whether 
any effect analogous to 
the dynamical Kondo effect
underlies the (nearly) continuous nature of the quantum phase transitions in those models.

{\it Summary. }
The $SU(2)$-symmetric Anderson lattice model has been studied using the EDMFT method,
and is shown to display a continuous AFM quantum phase transition.
We have established the Kondo-destruction nature of the QCP, demonstrated that the spin dynamics
obey $\omega/T$ scaling and found fractional temperature and frequency exponents.
Finally, we have reported the first calculation of the local entanglement entropy across the
Kondo-destruction QCP in any Anderson/Kondo lattice model. The result implies a dynamical Kondo effect, 
which is crucial for realizing the continuous nature of the quantum phase transition
and for elucidating the unusual properties of the quantum critical heavy fermion systems.
As such, our results
considerably  deepen the understanding of quantum critical
heavy fermion metals with continuous spin symmetry, and set the stage to link 
the beyond-Landau quantum criticality of heavy fermion systems with its counterpart in Mott-Hubbard systems.

We thank Kevin Ingersent, Stefan Kirchner, Chia-Chuan Liu, Silke Paschen, Jed Pixley, and Frank Steglich 
for useful discussions. The work was in part supported by the NSF (DMR-1920740)
 and the Robert A. Welch Foundation (C-1411).
Computing resources were supported in part by the
 Data Analysis and Visualization 
Cyberinfrastructure funded by NSF under grant OCI-0959097 and an IBM Shared University 
Research (SUR) Award at Rice University, and by the Extreme Science and Engineering
Discovery Environment (XSEDE) by NSF under Grants No. DMR170109.
Q.S.\ acknowledges 
the hospitality of the Aspen
Center for Physics (NSF, PHY-1607611).

\bibliography{edmft}

\begin{thebibliography}{47}
\expandafter\ifx\csname natexlab\endcsname\relax\def\natexlab#1{#1}\fi
\expandafter\ifx\csname bibnamefont\endcsname\relax
  \def\bibnamefont#1{#1}\fi
\expandafter\ifx\csname bibfnamefont\endcsname\relax
  \def\bibfnamefont#1{#1}\fi
\expandafter\ifx\csname citenamefont\endcsname\relax
  \def\citenamefont#1{#1}\fi
\expandafter\ifx\csname url\endcsname\relax
  \def\url#1{\texttt{#1}}\fi
\expandafter\ifx\csname urlprefix\endcsname\relax\def\urlprefix{URL }\fi
\providecommand{\bibinfo}[2]{#2}
\providecommand{\eprint}[2][]{\url{#2}}

\bibitem[{\citenamefont{{S}pecial issue: {Q}uantum~{P}hase
  {T}ransitions}(2010)}]{JLTP2010}
\bibinfo{author}{\bibnamefont{{S}pecial issue: {Q}uantum~{P}hase
  {T}ransitions}}, \bibinfo{journal}{J. Low Temp. Phys.}
  \textbf{\bibinfo{volume}{161}}, \bibinfo{pages}{1} (\bibinfo{year}{2010}).

\bibitem[{\citenamefont{Kirchner et~al.}(2020)\citenamefont{Kirchner, Paschen,
  Chen, Wirth, Feng, Thompson, and Si}}]{Kirchner_rmp20}
\bibinfo{author}{\bibfnamefont{S.}~\bibnamefont{Kirchner}},
  \bibinfo{author}{\bibfnamefont{S.}~\bibnamefont{Paschen}},
  \bibinfo{author}{\bibfnamefont{Q.}~\bibnamefont{Chen}},
  \bibinfo{author}{\bibfnamefont{S.}~\bibnamefont{Wirth}},
  \bibinfo{author}{\bibfnamefont{D.}~\bibnamefont{Feng}},
  \bibinfo{author}{\bibfnamefont{J.~D.} \bibnamefont{Thompson}},
  \bibnamefont{and} \bibinfo{author}{\bibfnamefont{Q.}~\bibnamefont{Si}},
  \bibinfo{journal}{Rev. Mod. Phys.} \textbf{\bibinfo{volume}{92}},
  \bibinfo{pages}{011002} (\bibinfo{year}{2020}).

\bibitem[{\citenamefont{Keimer and Moore}(2017)}]{KeimerMoore}
\bibinfo{author}{\bibfnamefont{B.}~\bibnamefont{Keimer}} \bibnamefont{and}
  \bibinfo{author}{\bibfnamefont{J.~E.} \bibnamefont{Moore}},
  \bibinfo{journal}{Nat. Phys.} \textbf{\bibinfo{volume}{13}},
  \bibinfo{pages}{1045} (\bibinfo{year}{2017}).

\bibitem[{\citenamefont{Coleman and Schofield}(2005)}]{Coleman-Nature}
\bibinfo{author}{\bibfnamefont{P.}~\bibnamefont{Coleman}} \bibnamefont{and}
  \bibinfo{author}{\bibfnamefont{A.~J.} \bibnamefont{Schofield}},
  \bibinfo{journal}{Nature} \textbf{\bibinfo{volume}{433}},
  \bibinfo{pages}{226} (\bibinfo{year}{2005}).

\bibitem[{\citenamefont{Sachdev}(1999)}]{Sachdev-book}
\bibinfo{author}{\bibfnamefont{S.}~\bibnamefont{Sachdev}},
  \emph{\bibinfo{title}{Quantum Phase Transitions}}
  (\bibinfo{publisher}{Cambridge University Press},
  \bibinfo{address}{Cambridge}, \bibinfo{year}{1999}).

\bibitem[{\citenamefont{Si et~al.}(2001)\citenamefont{Si, Rabello, Ingersent,
  and Smith}}]{Si-Nature}
\bibinfo{author}{\bibfnamefont{Q.}~\bibnamefont{Si}},
  \bibinfo{author}{\bibfnamefont{S.}~\bibnamefont{Rabello}},
  \bibinfo{author}{\bibfnamefont{K.}~\bibnamefont{Ingersent}},
  \bibnamefont{and} \bibinfo{author}{\bibfnamefont{J.}~\bibnamefont{Smith}},
  \bibinfo{journal}{Nature} \textbf{\bibinfo{volume}{413}},
  \bibinfo{pages}{804} (\bibinfo{year}{2001}).

\bibitem[{\citenamefont{Coleman et~al.}(2001)\citenamefont{Coleman, P\'{e}pin,
  Si, and Ramazashvili}}]{Colemanetal}
\bibinfo{author}{\bibfnamefont{P.}~\bibnamefont{Coleman}},
  \bibinfo{author}{\bibfnamefont{C.}~\bibnamefont{P\'{e}pin}},
  \bibinfo{author}{\bibfnamefont{Q.}~\bibnamefont{Si}}, \bibnamefont{and}
  \bibinfo{author}{\bibfnamefont{R.}~\bibnamefont{Ramazashvili}},
  \bibinfo{journal}{J.~Phys.~Cond.~Matt.} \textbf{\bibinfo{volume}{13}},
  \bibinfo{pages}{R723} (\bibinfo{year}{2001}).

\bibitem[{\citenamefont{Prochaska et~al.}(2020)\citenamefont{Prochaska, Li,
  MacFarland, Andrews, Bonta, Bianco, Yazdi, Schrenk, Detz, Limbeck
  et~al.}}]{Prochaska20}
\bibinfo{author}{\bibfnamefont{L.}~\bibnamefont{Prochaska}},
  \bibinfo{author}{\bibfnamefont{X.}~\bibnamefont{Li}},
  \bibinfo{author}{\bibfnamefont{D.~C.} \bibnamefont{MacFarland}},
  \bibinfo{author}{\bibfnamefont{A.~M.} \bibnamefont{Andrews}},
  \bibinfo{author}{\bibfnamefont{M.}~\bibnamefont{Bonta}},
  \bibinfo{author}{\bibfnamefont{E.~F.} \bibnamefont{Bianco}},
  \bibinfo{author}{\bibfnamefont{S.}~\bibnamefont{Yazdi}},
  \bibinfo{author}{\bibfnamefont{W.}~\bibnamefont{Schrenk}},
  \bibinfo{author}{\bibfnamefont{H.}~\bibnamefont{Detz}},
  \bibinfo{author}{\bibfnamefont{A.}~\bibnamefont{Limbeck}},
  \bibnamefont{et~al.}, \bibinfo{journal}{Science}
  \textbf{\bibinfo{volume}{367}}, \bibinfo{pages}{285} (\bibinfo{year}{2020}).

\bibitem[{\citenamefont{Schr\"oder et~al.}(2000)\citenamefont{Schr\"oder,
  Aeppli, Coldea, Adams, Stockert, v.~L\"ohneysen, Bucher, Ramazashvili, and
  Coleman}}]{schroder}
\bibinfo{author}{\bibfnamefont{A.}~\bibnamefont{Schr\"oder}},
  \bibinfo{author}{\bibfnamefont{G.}~\bibnamefont{Aeppli}},
  \bibinfo{author}{\bibfnamefont{R.}~\bibnamefont{Coldea}},
  \bibinfo{author}{\bibfnamefont{M.}~\bibnamefont{Adams}},
  \bibinfo{author}{\bibfnamefont{O.}~\bibnamefont{Stockert}},
  \bibinfo{author}{\bibfnamefont{H.}~\bibnamefont{v.~L\"ohneysen}},
  \bibinfo{author}{\bibfnamefont{E.}~\bibnamefont{Bucher}},
  \bibinfo{author}{\bibfnamefont{R.}~\bibnamefont{Ramazashvili}},
  \bibnamefont{and} \bibinfo{author}{\bibfnamefont{P.}~\bibnamefont{Coleman}},
  \bibinfo{journal}{Nature} \textbf{\bibinfo{volume}{407}},
  \bibinfo{pages}{351} (\bibinfo{year}{2000}).

\bibitem[{\citenamefont{Paschen et~al.}(2004)\citenamefont{Paschen,
  L{\"u}hmann, Wirth, Gegenwart, Trovarelli, Geibel, Steglich, Coleman, and
  Si}}]{paschen2004}
\bibinfo{author}{\bibfnamefont{S.}~\bibnamefont{Paschen}},
  \bibinfo{author}{\bibfnamefont{T.}~\bibnamefont{L{\"u}hmann}},
  \bibinfo{author}{\bibfnamefont{S.}~\bibnamefont{Wirth}},
  \bibinfo{author}{\bibfnamefont{P.}~\bibnamefont{Gegenwart}},
  \bibinfo{author}{\bibfnamefont{O.}~\bibnamefont{Trovarelli}},
  \bibinfo{author}{\bibfnamefont{C.}~\bibnamefont{Geibel}},
  \bibinfo{author}{\bibfnamefont{F.}~\bibnamefont{Steglich}},
  \bibinfo{author}{\bibfnamefont{P.}~\bibnamefont{Coleman}}, \bibnamefont{and}
  \bibinfo{author}{\bibfnamefont{Q.}~\bibnamefont{Si}},
  \bibinfo{journal}{Nature} \textbf{\bibinfo{volume}{432}},
  \bibinfo{pages}{881} (\bibinfo{year}{2004}).

\bibitem[{\citenamefont{Shishido et~al.}(2005)\citenamefont{Shishido, Settai,
  Harima, and \={O}nuki}}]{shishido2005}
\bibinfo{author}{\bibfnamefont{H.}~\bibnamefont{Shishido}},
  \bibinfo{author}{\bibfnamefont{R.}~\bibnamefont{Settai}},
  \bibinfo{author}{\bibfnamefont{H.}~\bibnamefont{Harima}}, \bibnamefont{and}
  \bibinfo{author}{\bibfnamefont{Y.}~\bibnamefont{\={O}nuki}},
  \bibinfo{journal}{J.~Phys.~Soc.~Jpn.} \textbf{\bibinfo{volume}{74}},
  \bibinfo{pages}{1103} (\bibinfo{year}{2005}).

\bibitem[{\citenamefont{Park et~al.}(2006)\citenamefont{Park, Ronning, Yuan,
  Salamon, Movshovich, Sarrao, and Thompson}}]{Par06.1}
\bibinfo{author}{\bibfnamefont{T.}~\bibnamefont{Park}},
  \bibinfo{author}{\bibfnamefont{F.}~\bibnamefont{Ronning}},
  \bibinfo{author}{\bibfnamefont{H.~Q.} \bibnamefont{Yuan}},
  \bibinfo{author}{\bibfnamefont{M.~B.} \bibnamefont{Salamon}},
  \bibinfo{author}{\bibfnamefont{R.}~\bibnamefont{Movshovich}},
  \bibinfo{author}{\bibfnamefont{J.~L.} \bibnamefont{Sarrao}},
  \bibnamefont{and} \bibinfo{author}{\bibfnamefont{J.~D.}
  \bibnamefont{Thompson}}, \bibinfo{journal}{Nature}
  \textbf{\bibinfo{volume}{440}}, \bibinfo{pages}{65} (\bibinfo{year}{2006}).

\bibitem[{\citenamefont{Senthil}(2008)}]{Senthil08.1}
\bibinfo{author}{\bibfnamefont{T.}~\bibnamefont{Senthil}},
  \bibinfo{journal}{Phys. Rev. B} \textbf{\bibinfo{volume}{78}},
  \bibinfo{pages}{035103} (\bibinfo{year}{2008}).

\bibitem[{\citenamefont{Terletska et~al.}(2011)\citenamefont{Terletska,
  Vu\ifmmode \check{c}\else \v{c}\fi{}i\ifmmode \check{c}\else
  \v{c}\fi{}evi\ifmmode~\acute{c}\else \'{c}\fi{},
  Tanaskovi\ifmmode~\acute{c}\else \'{c}\fi{}, and
  Dobrosavljevi\ifmmode~\acute{c}\else \'{c}\fi{}}}]{Ter11}
\bibinfo{author}{\bibfnamefont{H.}~\bibnamefont{Terletska}},
  \bibinfo{author}{\bibfnamefont{J.}~\bibnamefont{Vu\ifmmode \check{c}\else
  \v{c}\fi{}i\ifmmode \check{c}\else \v{c}\fi{}evi\ifmmode~\acute{c}\else
  \'{c}\fi{}}},
  \bibinfo{author}{\bibfnamefont{D.}~\bibnamefont{Tanaskovi\ifmmode~\acute{c}\else
  \'{c}\fi{}}}, \bibnamefont{and}
  \bibinfo{author}{\bibfnamefont{V.}~\bibnamefont{Dobrosavljevi\ifmmode~\acute{c}\else
  \'{c}\fi{}}}, \bibinfo{journal}{Phys. Rev. Lett.}
  \textbf{\bibinfo{volume}{107}}, \bibinfo{pages}{026401}
  (\bibinfo{year}{2011}).

\bibitem[{\citenamefont{Ramshaw et~al.}(2015)\citenamefont{Ramshaw, Sebastian,
  McDonald, Day, Tan, Zhu, Betts, Liang, Bonn, Hardy et~al.}}]{Ram15.1}
\bibinfo{author}{\bibfnamefont{B.~J.} \bibnamefont{Ramshaw}},
  \bibinfo{author}{\bibfnamefont{S.~E.} \bibnamefont{Sebastian}},
  \bibinfo{author}{\bibfnamefont{R.~D.} \bibnamefont{McDonald}},
  \bibinfo{author}{\bibfnamefont{.~J.} \bibnamefont{Day}},
  \bibinfo{author}{\bibfnamefont{B.~S.} \bibnamefont{Tan}},
  \bibinfo{author}{\bibfnamefont{Z.}~\bibnamefont{Zhu}},
  \bibinfo{author}{\bibfnamefont{J.~B.} \bibnamefont{Betts}},
  \bibinfo{author}{\bibfnamefont{R.}~\bibnamefont{Liang}},
  \bibinfo{author}{\bibfnamefont{D.~A.} \bibnamefont{Bonn}},
  \bibinfo{author}{\bibfnamefont{W.~N.} \bibnamefont{Hardy}},
  \bibnamefont{et~al.}, \bibinfo{journal}{Science}
  \textbf{\bibinfo{volume}{348}}, \bibinfo{pages}{317} (\bibinfo{year}{2015}).

\bibitem[{\citenamefont{Badoux et~al.}(2016)\citenamefont{Badoux, Tabis,
  Lalibert\'e, Grissonnanche, Vignolle, Vignolles, B\'eard, Bonn, Hardy, Liang
  et~al.}}]{Bad16.1}
\bibinfo{author}{\bibfnamefont{S.}~\bibnamefont{Badoux}},
  \bibinfo{author}{\bibfnamefont{W.}~\bibnamefont{Tabis}},
  \bibinfo{author}{\bibfnamefont{F.}~\bibnamefont{Lalibert\'e}},
  \bibinfo{author}{\bibfnamefont{G.~.} \bibnamefont{Grissonnanche}},
  \bibinfo{author}{\bibfnamefont{B.}~\bibnamefont{Vignolle}},
  \bibinfo{author}{\bibfnamefont{D.}~\bibnamefont{Vignolles}},
  \bibinfo{author}{\bibfnamefont{J.}~\bibnamefont{B\'eard}},
  \bibinfo{author}{\bibfnamefont{D.~A.} \bibnamefont{Bonn}},
  \bibinfo{author}{\bibfnamefont{W.~N.} \bibnamefont{Hardy}},
  \bibinfo{author}{\bibfnamefont{R.}~\bibnamefont{Liang}},
  \bibnamefont{et~al.}, \bibinfo{journal}{{Nature}}
  \textbf{\bibinfo{volume}{531}}, \bibinfo{pages}{210} (\bibinfo{year}{2016}).

\bibitem[{\citenamefont{Si et~al.}(2003)\citenamefont{Si, Rabello, Ingersent,
  and Smith}}]{local_fluct_qm_2003}
\bibinfo{author}{\bibfnamefont{Q.}~\bibnamefont{Si}},
  \bibinfo{author}{\bibfnamefont{S.}~\bibnamefont{Rabello}},
  \bibinfo{author}{\bibfnamefont{K.}~\bibnamefont{Ingersent}},
  \bibnamefont{and} \bibinfo{author}{\bibfnamefont{J.~L.} \bibnamefont{Smith}},
  \bibinfo{journal}{Phys. Rev. B} \textbf{\bibinfo{volume}{68}},
  \bibinfo{pages}{115103} (\bibinfo{year}{2003}).

\bibitem[{\citenamefont{Si et~al.}(2014)\citenamefont{Si, Pixley, Nica,
  Yamamoto, Goswami, Yu, and Kirchner}}]{si2014kondo}
\bibinfo{author}{\bibfnamefont{Q.}~\bibnamefont{Si}},
  \bibinfo{author}{\bibfnamefont{J.~H.} \bibnamefont{Pixley}},
  \bibinfo{author}{\bibfnamefont{E.}~\bibnamefont{Nica}},
  \bibinfo{author}{\bibfnamefont{S.~J.} \bibnamefont{Yamamoto}},
  \bibinfo{author}{\bibfnamefont{P.}~\bibnamefont{Goswami}},
  \bibinfo{author}{\bibfnamefont{R.}~\bibnamefont{Yu}}, \bibnamefont{and}
  \bibinfo{author}{\bibfnamefont{S.}~\bibnamefont{Kirchner}},
  \bibinfo{journal}{Journal of the Physical Society of Japan}
  \textbf{\bibinfo{volume}{83}}, \bibinfo{pages}{061005}
  (\bibinfo{year}{2014}).

\bibitem[{\citenamefont{Zhu et~al.}(2003)\citenamefont{Zhu, Grempel, and
  Si}}]{jxzhu2003}
\bibinfo{author}{\bibfnamefont{J.-X.} \bibnamefont{Zhu}},
  \bibinfo{author}{\bibfnamefont{D.~R.} \bibnamefont{Grempel}},
  \bibnamefont{and} \bibinfo{author}{\bibfnamefont{Q.}~\bibnamefont{Si}},
  \bibinfo{journal}{Phys. Rev. Lett.} \textbf{\bibinfo{volume}{91}},
  \bibinfo{pages}{156404} (\bibinfo{year}{2003}).

\bibitem[{\citenamefont{Glossop and Ingersent}(2007)}]{glossop2007prl}
\bibinfo{author}{\bibfnamefont{M.~T.} \bibnamefont{Glossop}} \bibnamefont{and}
  \bibinfo{author}{\bibfnamefont{K.}~\bibnamefont{Ingersent}},
  \bibinfo{journal}{Phys. Rev. Lett.} \textbf{\bibinfo{volume}{99}},
  \bibinfo{pages}{227203} (\bibinfo{year}{2007}).

\bibitem[{\citenamefont{Zhu et~al.}(2007)\citenamefont{Zhu, Kirchner, Bulla,
  and Si}}]{jxzhu2007}
\bibinfo{author}{\bibfnamefont{J.-X.} \bibnamefont{Zhu}},
  \bibinfo{author}{\bibfnamefont{S.}~\bibnamefont{Kirchner}},
  \bibinfo{author}{\bibfnamefont{R.}~\bibnamefont{Bulla}}, \bibnamefont{and}
  \bibinfo{author}{\bibfnamefont{Q.}~\bibnamefont{Si}}, \bibinfo{journal}{Phys.
  Rev. Lett.} \textbf{\bibinfo{volume}{99}}, \bibinfo{pages}{227204}
  (\bibinfo{year}{2007}).

\bibitem[{\citenamefont{Das et~al.}(2014)\citenamefont{Das, Lin, Ghimire,
  Huang, Ronning, Bauer, Thompson, Batista, Ehlers, and Janoschek}}]{Das_prl14}
\bibinfo{author}{\bibfnamefont{P.}~\bibnamefont{Das}},
  \bibinfo{author}{\bibfnamefont{S.-Z.} \bibnamefont{Lin}},
  \bibinfo{author}{\bibfnamefont{N.~J.} \bibnamefont{Ghimire}},
  \bibinfo{author}{\bibfnamefont{K.}~\bibnamefont{Huang}},
  \bibinfo{author}{\bibfnamefont{F.}~\bibnamefont{Ronning}},
  \bibinfo{author}{\bibfnamefont{E.~D.} \bibnamefont{Bauer}},
  \bibinfo{author}{\bibfnamefont{J.~D.} \bibnamefont{Thompson}},
  \bibinfo{author}{\bibfnamefont{C.~D.} \bibnamefont{Batista}},
  \bibinfo{author}{\bibfnamefont{G.}~\bibnamefont{Ehlers}}, \bibnamefont{and}
  \bibinfo{author}{\bibfnamefont{M.}~\bibnamefont{Janoschek}},
  \bibinfo{journal}{Phys. Rev. Lett.} \textbf{\bibinfo{volume}{113}},
  \bibinfo{pages}{246403} (\bibinfo{year}{2014}).

\bibitem[{\citenamefont{Si and Smith}(1996)}]{si1996kosterlitz}
\bibinfo{author}{\bibfnamefont{Q.}~\bibnamefont{Si}} \bibnamefont{and}
  \bibinfo{author}{\bibfnamefont{J.~L.} \bibnamefont{Smith}},
  \bibinfo{journal}{Phys. Rev. Lett.} \textbf{\bibinfo{volume}{77}},
  \bibinfo{pages}{3391} (\bibinfo{year}{1996}).

\bibitem[{\citenamefont{Smith and Si}(2000)}]{Smith_edmft}
\bibinfo{author}{\bibfnamefont{J.~L.} \bibnamefont{Smith}} \bibnamefont{and}
  \bibinfo{author}{\bibfnamefont{Q.}~\bibnamefont{Si}}, \bibinfo{journal}{Phys.
  Rev. B} \textbf{\bibinfo{volume}{61}}, \bibinfo{pages}{5184}
  (\bibinfo{year}{2000}).

\bibitem[{\citenamefont{Chitra and Kotliar}(2000)}]{chitra2000effect}
\bibinfo{author}{\bibfnamefont{R.}~\bibnamefont{Chitra}} \bibnamefont{and}
  \bibinfo{author}{\bibfnamefont{G.}~\bibnamefont{Kotliar}},
  \bibinfo{journal}{Phys. Rev. Lett.} \textbf{\bibinfo{volume}{84}},
  \bibinfo{pages}{3678} (\bibinfo{year}{2000}).

\bibitem[{\citenamefont{Cai and Si}(2019)}]{Ang_su2_bf}
\bibinfo{author}{\bibfnamefont{A.}~\bibnamefont{Cai}} \bibnamefont{and}
  \bibinfo{author}{\bibfnamefont{Q.}~\bibnamefont{Si}}, \bibinfo{journal}{Phys.
  Rev. B} \textbf{\bibinfo{volume}{100}}, \bibinfo{pages}{014439}
  (\bibinfo{year}{2019}).

\bibitem[{\citenamefont{Gull et~al.}(2011)\citenamefont{Gull, Millis,
  Lichtenstein, Rubtsov, Troyer, and Werner}}]{Gull11}
\bibinfo{author}{\bibfnamefont{E.}~\bibnamefont{Gull}},
  \bibinfo{author}{\bibfnamefont{A.~J.} \bibnamefont{Millis}},
  \bibinfo{author}{\bibfnamefont{A.~I.} \bibnamefont{Lichtenstein}},
  \bibinfo{author}{\bibfnamefont{A.~N.} \bibnamefont{Rubtsov}},
  \bibinfo{author}{\bibfnamefont{M.}~\bibnamefont{Troyer}}, \bibnamefont{and}
  \bibinfo{author}{\bibfnamefont{P.}~\bibnamefont{Werner}},
  \bibinfo{journal}{Rev. Mod. Phys.} \textbf{\bibinfo{volume}{83}},
  \bibinfo{pages}{349} (\bibinfo{year}{2011}).

\bibitem[{\citenamefont{Otsuki}(2013)}]{Otsuki_su2_bf}
\bibinfo{author}{\bibfnamefont{J.}~\bibnamefont{Otsuki}},
  \bibinfo{journal}{Phys. Rev. B} \textbf{\bibinfo{volume}{87}},
  \bibinfo{pages}{125102} (\bibinfo{year}{2013}).

\bibitem[{\citenamefont{Pixley et~al.}(2015)\citenamefont{Pixley, Chowdhury,
  Miecnikowski, Stephens, Wagner, and Ingersent}}]{EE_KD}
\bibinfo{author}{\bibfnamefont{J.~H.} \bibnamefont{Pixley}},
  \bibinfo{author}{\bibfnamefont{T.}~\bibnamefont{Chowdhury}},
  \bibinfo{author}{\bibfnamefont{M.~T.} \bibnamefont{Miecnikowski}},
  \bibinfo{author}{\bibfnamefont{J.}~\bibnamefont{Stephens}},
  \bibinfo{author}{\bibfnamefont{C.}~\bibnamefont{Wagner}}, \bibnamefont{and}
  \bibinfo{author}{\bibfnamefont{K.}~\bibnamefont{Ingersent}},
  \bibinfo{journal}{Phys. Rev. B} \textbf{\bibinfo{volume}{91}},
  \bibinfo{pages}{245122} (\bibinfo{year}{2015}).

\bibitem[{\citenamefont{Wagner et~al.}(2018)\citenamefont{Wagner, Chowdhury,
  Pixley, and Ingersent}}]{EE_KD_long_range}
\bibinfo{author}{\bibfnamefont{C.}~\bibnamefont{Wagner}},
  \bibinfo{author}{\bibfnamefont{T.}~\bibnamefont{Chowdhury}},
  \bibinfo{author}{\bibfnamefont{J.~H.} \bibnamefont{Pixley}},
  \bibnamefont{and}
  \bibinfo{author}{\bibfnamefont{K.}~\bibnamefont{Ingersent}},
  \bibinfo{journal}{Phys. Rev. Lett.} \textbf{\bibinfo{volume}{121}},
  \bibinfo{pages}{147602} (\bibinfo{year}{2018}).

\bibitem[{sup()}]{supp}
\emph{\bibinfo{title}{Supplemental material}}.

\bibitem[{\citenamefont{Si et~al.}(2005)\citenamefont{Si, Zhu, and
  Grempel}}]{Si_2005}
\bibinfo{author}{\bibfnamefont{Q.}~\bibnamefont{Si}},
  \bibinfo{author}{\bibfnamefont{J.-X.} \bibnamefont{Zhu}}, \bibnamefont{and}
  \bibinfo{author}{\bibfnamefont{D.~R.} \bibnamefont{Grempel}},
  \bibinfo{journal}{Journal of Physics: Condensed Matter}
  \textbf{\bibinfo{volume}{17}}, \bibinfo{pages}{R1025} (\bibinfo{year}{2005}).

\bibitem[{\citenamefont{Smith and Si}(1999)}]{smith1999}
\bibinfo{author}{\bibfnamefont{J.~L.} \bibnamefont{Smith}} \bibnamefont{and}
  \bibinfo{author}{\bibfnamefont{Q.}~\bibnamefont{Si}},
  \bibinfo{journal}{Europhysics Letters ({EPL})} \textbf{\bibinfo{volume}{45}},
  \bibinfo{pages}{228} (\bibinfo{year}{1999}).

\bibitem[{\citenamefont{Sengupta}(2000)}]{sengupta2000}
\bibinfo{author}{\bibfnamefont{A.~M.} \bibnamefont{Sengupta}},
  \bibinfo{journal}{Phys. Rev. B} \textbf{\bibinfo{volume}{61}},
  \bibinfo{pages}{4041} (\bibinfo{year}{2000}).

\bibitem[{\citenamefont{Zhu and Si}(2002)}]{zhu2002}
\bibinfo{author}{\bibfnamefont{L.}~\bibnamefont{Zhu}} \bibnamefont{and}
  \bibinfo{author}{\bibfnamefont{Q.}~\bibnamefont{Si}}, \bibinfo{journal}{Phys.
  Rev. B} \textbf{\bibinfo{volume}{66}}, \bibinfo{pages}{024426}
  (\bibinfo{year}{2002}).

\bibitem[{\citenamefont{Zar{\'a}nd and Demler}(2002)}]{zarand2002}
\bibinfo{author}{\bibfnamefont{G.}~\bibnamefont{Zar{\'a}nd}} \bibnamefont{and}
  \bibinfo{author}{\bibfnamefont{E.}~\bibnamefont{Demler}},
  \bibinfo{journal}{Phys. Rev. B} \textbf{\bibinfo{volume}{66}},
  \bibinfo{pages}{024427} (\bibinfo{year}{2002}).

\bibitem[{\citenamefont{Grempel and Si}(2003)}]{Gre03.2}
\bibinfo{author}{\bibfnamefont{D.}~\bibnamefont{Grempel}} \bibnamefont{and}
  \bibinfo{author}{\bibfnamefont{Q.}~\bibnamefont{Si}},
  \bibinfo{journal}{{Phys.\ Rev.\ Lett.}} \textbf{\bibinfo{volume}{91}},
  \bibinfo{pages}{026401} (\bibinfo{year}{2003}).

\bibitem[{\citenamefont{Hertz}(1976)}]{hertz1976quantum}
\bibinfo{author}{\bibfnamefont{J.~A.} \bibnamefont{Hertz}},
  \bibinfo{journal}{Phys. Rev. B} \textbf{\bibinfo{volume}{14}},
  \bibinfo{pages}{1165} (\bibinfo{year}{1976}).

\bibitem[{\citenamefont{Millis}(1993)}]{millis1993effect}
\bibinfo{author}{\bibfnamefont{A.}~\bibnamefont{Millis}},
  \bibinfo{journal}{Phys. Rev. B} \textbf{\bibinfo{volume}{48}},
  \bibinfo{pages}{7183} (\bibinfo{year}{1993}).

\bibitem[{\citenamefont{Moriya}(2012)}]{moriya2012spin}
\bibinfo{author}{\bibfnamefont{T.}~\bibnamefont{Moriya}},
  \emph{\bibinfo{title}{Spin fluctuations in itinerant electron magnetism}},
  vol.~\bibinfo{volume}{56} (\bibinfo{publisher}{Springer Science \& Business
  Media}, \bibinfo{year}{2012}).

\bibitem[{\citenamefont{Larsson and Johannesson}(2006)}]{EE_loc}
\bibinfo{author}{\bibfnamefont{D.}~\bibnamefont{Larsson}} \bibnamefont{and}
  \bibinfo{author}{\bibfnamefont{H.}~\bibnamefont{Johannesson}},
  \bibinfo{journal}{Phys. Rev. A} \textbf{\bibinfo{volume}{73}},
  \bibinfo{pages}{042320} (\bibinfo{year}{2006}).

\bibitem[{\citenamefont{Cai et~al.}(2019)\citenamefont{Cai, Hu, Ingersent,
  Paschen, and Si}}]{dynamical_kondo}
\bibinfo{author}{\bibfnamefont{A.}~\bibnamefont{Cai}},
  \bibinfo{author}{\bibfnamefont{H.}~\bibnamefont{Hu}},
  \bibinfo{author}{\bibfnamefont{K.}~\bibnamefont{Ingersent}},
  \bibinfo{author}{\bibfnamefont{S.}~\bibnamefont{Paschen}}, \bibnamefont{and}
  \bibinfo{author}{\bibfnamefont{Q.}~\bibnamefont{Si}},
  \bibinfo{journal}{arXiv:1904.11471}  (\bibinfo{year}{2019}).

\bibitem[{\citenamefont{Martelli et~al.}(2019)\citenamefont{Martelli, Cai,
  Nica, Taupin, Prokofiev, Liu, Lai, Yu, Ingersent, K{\"u}chler
  et~al.}}]{Mar18.1}
\bibinfo{author}{\bibfnamefont{V.}~\bibnamefont{Martelli}},
  \bibinfo{author}{\bibfnamefont{A.}~\bibnamefont{Cai}},
  \bibinfo{author}{\bibfnamefont{E.~M.} \bibnamefont{Nica}},
  \bibinfo{author}{\bibfnamefont{M.}~\bibnamefont{Taupin}},
  \bibinfo{author}{\bibfnamefont{A.}~\bibnamefont{Prokofiev}},
  \bibinfo{author}{\bibfnamefont{C.-C.} \bibnamefont{Liu}},
  \bibinfo{author}{\bibfnamefont{H.-H.} \bibnamefont{Lai}},
  \bibinfo{author}{\bibfnamefont{R.}~\bibnamefont{Yu}},
  \bibinfo{author}{\bibfnamefont{K.}~\bibnamefont{Ingersent}},
  \bibinfo{author}{\bibfnamefont{R.}~\bibnamefont{K{\"u}chler}},
  \bibnamefont{et~al.}, \bibinfo{journal}{Proceedings of the National Academy
  of Sciences} \textbf{\bibinfo{volume}{116}}, \bibinfo{pages}{17701}
  (\bibinfo{year}{2019}).

\bibitem[{\citenamefont{Gegenwart et~al.}(2002)\citenamefont{Gegenwart,
  Custers, Geibel, Neumaier, Tayama, Tenya, Trovarelli, and
  Steglich}}]{Geg02.1}
\bibinfo{author}{\bibfnamefont{P.}~\bibnamefont{Gegenwart}},
  \bibinfo{author}{\bibfnamefont{J.}~\bibnamefont{Custers}},
  \bibinfo{author}{\bibfnamefont{C.}~\bibnamefont{Geibel}},
  \bibinfo{author}{\bibfnamefont{K.}~\bibnamefont{Neumaier}},
  \bibinfo{author}{\bibfnamefont{T.}~\bibnamefont{Tayama}},
  \bibinfo{author}{\bibfnamefont{K.}~\bibnamefont{Tenya}},
  \bibinfo{author}{\bibfnamefont{O.}~\bibnamefont{Trovarelli}},
  \bibnamefont{and} \bibinfo{author}{\bibfnamefont{F.}~\bibnamefont{Steglich}},
  \bibinfo{journal}{{Phys.\ Rev.\ Lett.}} \textbf{\bibinfo{volume}{89}},
  \bibinfo{pages}{056402} (\bibinfo{year}{2002}).

\bibitem[{\citenamefont{Joshi et~al.}(2020)\citenamefont{Joshi, Li,
  Tarnopolsky, Georges, and Sachdev}}]{Joshi20}
\bibinfo{author}{\bibfnamefont{D.}~\bibnamefont{Joshi}},
  \bibinfo{author}{\bibfnamefont{C.}~\bibnamefont{Li}},
  \bibinfo{author}{\bibfnamefont{G.}~\bibnamefont{Tarnopolsky}},
  \bibinfo{author}{\bibfnamefont{A.}~\bibnamefont{Georges}}, \bibnamefont{and}
  \bibinfo{author}{\bibfnamefont{S.}~\bibnamefont{Sachdev}},
  \bibinfo{journal}{arXiv:1912.08822}  (\bibinfo{year}{2020}).

\bibitem[{\citenamefont{Cha et~al.}(2020)\citenamefont{Cha, Wentzell,
  Parcollet, Georges, and Kim}}]{Cha20}
\bibinfo{author}{\bibfnamefont{P.}~\bibnamefont{Cha}},
  \bibinfo{author}{\bibfnamefont{N.}~\bibnamefont{Wentzell}},
  \bibinfo{author}{\bibfnamefont{O.}~\bibnamefont{Parcollet}},
  \bibinfo{author}{\bibfnamefont{A.}~\bibnamefont{Georges}}, \bibnamefont{and}
  \bibinfo{author}{\bibfnamefont{E.}~\bibnamefont{Kim}},
  \bibinfo{journal}{arXiv:2002.07181}  (\bibinfo{year}{2020}).

\bibitem[{\citenamefont{Tarnopolsky et~al.}(2020)\citenamefont{Tarnopolsky, Li,
  Joshi, and Sachdev}}]{Tarnopolsky20}
\bibinfo{author}{\bibfnamefont{G.}~\bibnamefont{Tarnopolsky}},
  \bibinfo{author}{\bibfnamefont{C.}~\bibnamefont{Li}},
  \bibinfo{author}{\bibfnamefont{D.}~\bibnamefont{Joshi}}, \bibnamefont{and}
  \bibinfo{author}{\bibfnamefont{S.}~\bibnamefont{Sachdev}},
  \bibinfo{journal}{arXiv:2002.12381}  (\bibinfo{year}{2020}).

\end{thebibliography}


\begin{thebibliography}{3}
\expandafter\ifx\csname natexlab\endcsname\relax\def\natexlab#1{#1}\fi
\expandafter\ifx\csname bibnamefont\endcsname\relax
  \def\bibnamefont#1{#1}\fi
\expandafter\ifx\csname bibfnamefont\endcsname\relax
  \def\bibfnamefont#1{#1}\fi
\expandafter\ifx\csname citenamefont\endcsname\relax
  \def\citenamefont#1{#1}\fi
\expandafter\ifx\csname url\endcsname\relax
  \def\url#1{\texttt{#1}}\fi
\expandafter\ifx\csname urlprefix\endcsname\relax\def\urlprefix{URL }\fi
\providecommand{\bibinfo}[2]{#2}
\providecommand{\eprint}[2][]{\url{#2}}

\bibitem[{\citenamefont{Zhu and Si}(2002)}]{zhu2002}
\bibinfo{author}{\bibfnamefont{L.}~\bibnamefont{Zhu}} \bibnamefont{and}
  \bibinfo{author}{\bibfnamefont{Q.}~\bibnamefont{Si}}, \bibinfo{journal}{Phys.
  Rev. B} \textbf{\bibinfo{volume}{66}}, \bibinfo{pages}{024426}
  (\bibinfo{year}{2002}).

\bibitem[{\citenamefont{Si and Smith}(1996)}]{si1996kosterlitz}
\bibinfo{author}{\bibfnamefont{Q.}~\bibnamefont{Si}} \bibnamefont{and}
  \bibinfo{author}{\bibfnamefont{J.~L.} \bibnamefont{Smith}},
  \bibinfo{journal}{Phys. Rev. Lett.} \textbf{\bibinfo{volume}{77}},
  \bibinfo{pages}{3391} (\bibinfo{year}{1996}).

\bibitem[{\citenamefont{Smith and Si}(2000)}]{Smith_edmft}
\bibinfo{author}{\bibfnamefont{J.~L.} \bibnamefont{Smith}} \bibnamefont{and}
  \bibinfo{author}{\bibfnamefont{Q.}~\bibnamefont{Si}}, \bibinfo{journal}{Phys.
  Rev. B} \textbf{\bibinfo{volume}{61}}, \bibinfo{pages}{5184}
  (\bibinfo{year}{2000}).

\end{thebibliography}

\end{document}


\title{Supplementary Material}
\maketitle

\onecolumngrid 

\section{Extended dynamical mean field theory}

The EDMFT approach treats the Anderson lattice model, Eq.\,(1) of the main text, in terms of a Bose-Fermi 
Anderson model:
\begin{eqnarray}
&H &= H_{loc} +H_{B} + H_{F} \nonumber\\
&H_{loc}&= 
\frac{U}{2}( n_{d,\uparrow}+n_{d,\downarrow}-1)^2 
+h_{loc} S^z
 \nonumber \\
&H_{B} &= \sum_{p} \omega_p \bm{\phi}_p^\dag \cdot 
\bm{\phi}_p +g \sum_{p} \bm{S}\cdot (\bm{\phi}_{p}^\dag +\bm{\phi}_{-p})
 \nonumber\\
&H_{F} &= \sum_{p,\sigma} E_p c_{p,\sigma}^\dag c_{p,\sigma} 
+V\sum_{p,\sigma} (c_{p,\sigma}^\dag d_{\sigma} +d_{\sigma}^\dag  c_{p,\sigma}  ) \, .
\end{eqnarray} 
Here, $n_{d,\sigma} = d_{\sigma}^{\dagger} d_{\sigma}$, and 
$c_{p,\sigma}$ describes the fermionic bath with dispersion $E_p$.
In addition, 
$\bm{\phi}_p$ represents a three-component bosonic bath, with dispersion $\omega_p$; it captures the fluctuations 
of the local moments. Integrating out the $\bm{\phi}$ bosons and $c$ fermions leads to the action given in Eq.\,2 of the main text.

\section{Identifying the quantum critical point}
\begin{figure}[b!]
\captionsetup[subfigure]{labelformat=empty}
  \centering
    \mbox{\includegraphics[width=0.65\columnwidth]{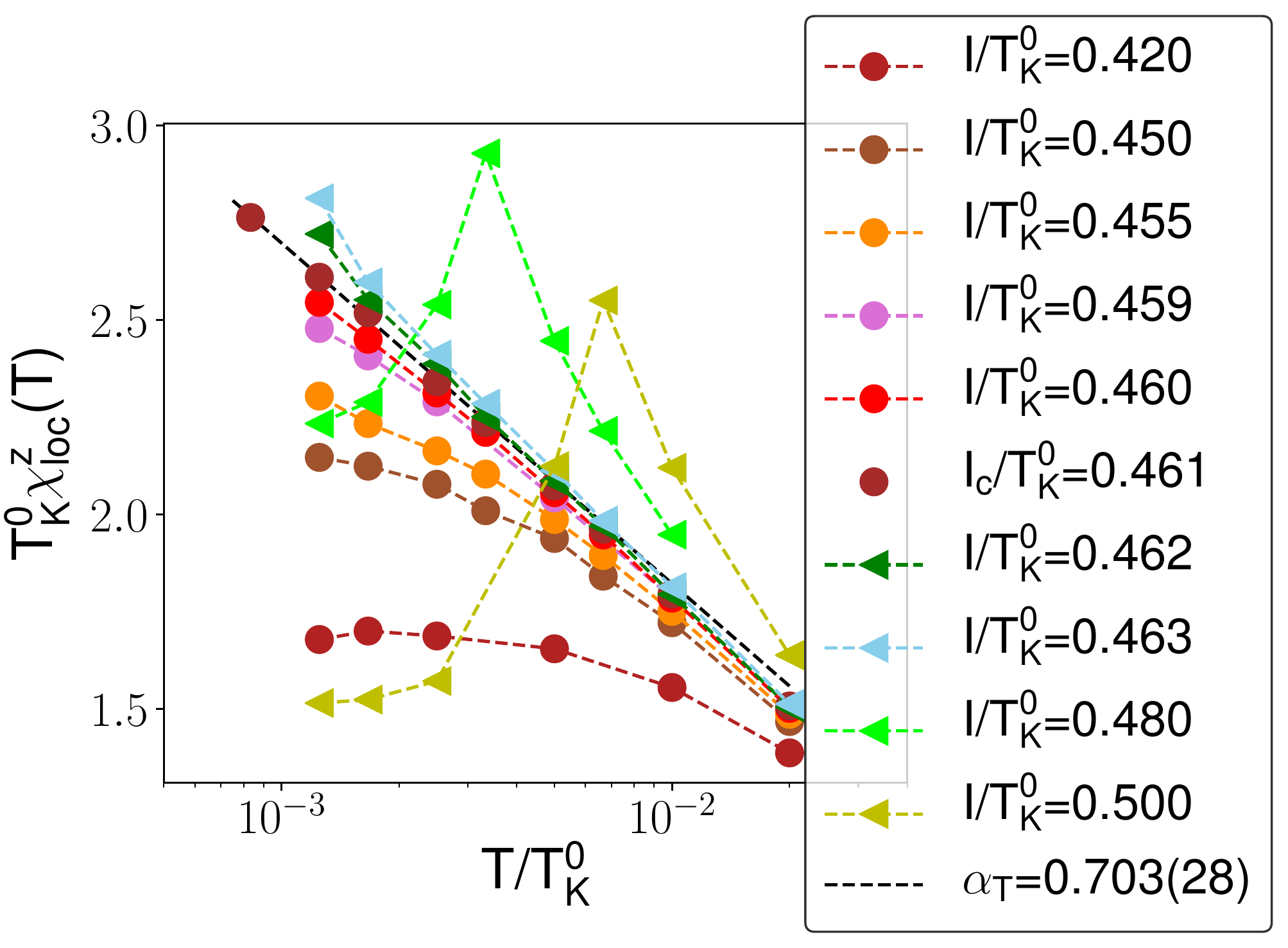}}    
\caption{(Color online) Static local susceptibility as a function of temperature, for a dense set of RKKY interactions $I$.  }
\label{fig:chi_loc_qcp}
\end{figure}
We identify the critical point through two complementary procedures. One is 
the extrapolation of the $T_{N}(I)$ curve to zero temperature. Another is to examine
the behavior of the static local susceptibility.
As we discussed in the main text, the static local susceptibility has different behavior in different phases. 
In the Kondo-screened phase, $\chi^z_{loc}(T)$ saturates when we lower the temperature $T$ towards zero.
In the AFM phase, as we lower the temperature,
$\chi^z_{loc}(T)$ first goes upwards, which corresponds to the Curie-Weiss behavior. 
It  then drops off upon the onset of the 
AFM order.
Based on this insight, we can search for the critical RKKY interaction $I_c$ (as mentioned in the main text, the bare Kondo scale is 
fixed) by the point where the singular (logarithmic) behavior survives to the lowest temperature.
We show the evolution of $\chi^z_{loc}(T)$
 at different ratios of $I/T_K^0$ in Fig.~\ref{fig:chi_loc_qcp}.
At $I=I_c=0.461T_K^0$, $\chi^z_{loc}(T)$ has the most straight-line behavior in the semi-log plot.

\section{Kondo destruction and the renormalization group flow of the Bose-Fermi Kondo model}
\begin{figure}[t!]
\captionsetup[subfigure]{labelformat=empty}
  \centering
    \mbox{\includegraphics[width=0.6\columnwidth]{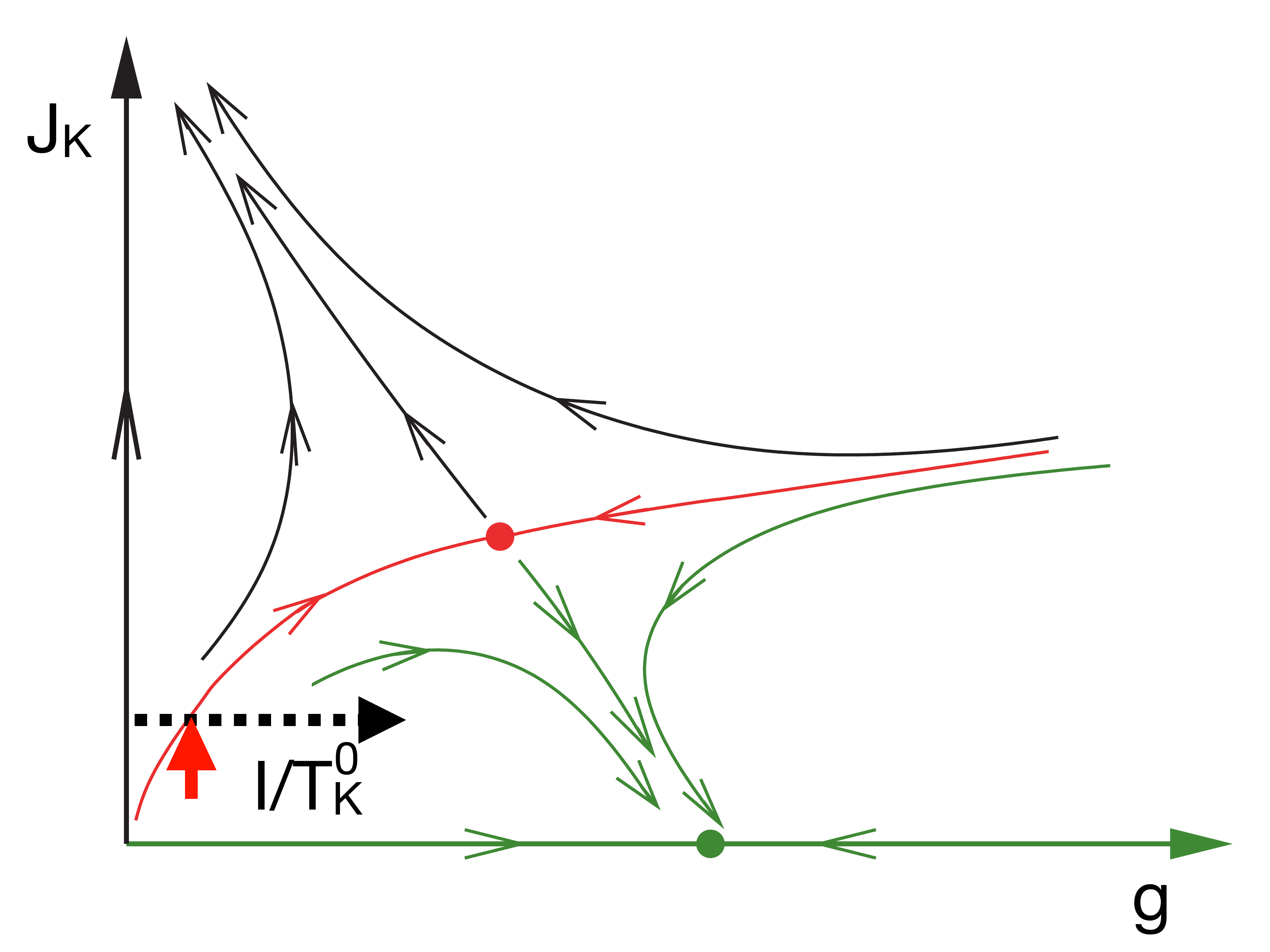}}    
\caption{(Color online) RG flow of the Bose-Fermi Kondo/Anderson model \cite{zhu2002}. 
$g$ is the coupling between the local moment and bosonic bath, and $J_K$ is the Kondo coupling between 
the local moment and fermionic bath.}
\label{fig:RG}
\end{figure}
In Fig.~\ref{fig:RG}, we show the renormalization group (RG) flow of the effective Bose-Fermi Kondo model \cite{zhu2002}, 
where $g$ denotes the coupling constant between the local moment and bosonic bath, and $J_K$ represents the Kondo coupling. 
In our study of the Anderson lattice model,
we fixed the bare Kondo temperature $T_K^0$ and tune the RKKY interaction $I$.
For the corresponding Bose-Fermi impurity model, this is illustrated by the dashed
arrow in Fig.~\ref{fig:RG}. 
The model may flow to three fixed points by tuning $I/T_K^0$. For small $I/T_K^0$, the impurity 
model flows to a Kondo-screened fixed point.
For large $I/T_K^0$, the model flows to a Kondo-destroyed fixed point with zero $J_K$. 
In between, the system passes through the separatrix (the red arrow in Fig.~\ref{fig:RG}), where it flows to the Kondo-destruction QCP.
On the separatrix, {\it e.g.} upon reaching the red arrow following
the dashed line from left in Fig.~\ref{fig:RG}, the local spin susceptibility $\chi_{loc}(T\rightarrow 0)$ diverges.
The observations from the RG flow are consistent with the results presented in the main text.

\section{Relation between the local susceptibility and antiferromagnetic lattice susceptibility}
In this section, we summarize the calculation of the antiferromagnetic lattice susceptibility,
which in the paramagnetic phase is determined by the Dyson equation\cite{si1996kosterlitz,Smith_edmft}:
\ba
 \chi^{\alpha} ({\bm q},i\omega_n)=\langle {S}^{\alpha}({\bm q},i\omega_n)  S^{\alpha}({\bm q},-i\omega_n)  \rangle
 =1/\left [ M^{\alpha}(i\omega_n) + I_{\bm q} \right ]\, , 
\ea
with $\alpha \in \{x,y,z\}$.
At the antiferromagnetic wave vector ${\bf Q}$, we have
\begin{equation}
\chi ^{\alpha}({\bf Q},i\omega_n) = 1/\left [ M^{\alpha}(i\omega_n)-2I \right ]  \,.
\label{eq:chi_lat}
\end{equation}
Combining the above equation with the self-consistent equations in the main text, 
we can derive the following asymptotic relation between $\chi_{loc}(i\omega_n)$ and $\chi ({\bf Q},i\omega_n)$, 
which is valid in the region of $4I\chi_{loc}(i\omega_n) \gg 1$
\begin{eqnarray}
\chi ^{\alpha}({\bf Q},i\omega_n) &=&\frac{1}{4I} [\exp(4I\chi^{\alpha}_{loc}(i\omega_n) )-1] \nonumber \\
&\approx &\frac{\exp(4I\chi^{\alpha}_{loc}(i\omega_n))}{4I} 
\label{eq:relation}
\end{eqnarray}.

\section{Frequency dependence of the antiferromagnetic dynamical spin susceptibility}
\begin{figure}
\captionsetup[subfigure]{labelformat=empty}
  \centering
    \mbox{\includegraphics[width=0.7\columnwidth]{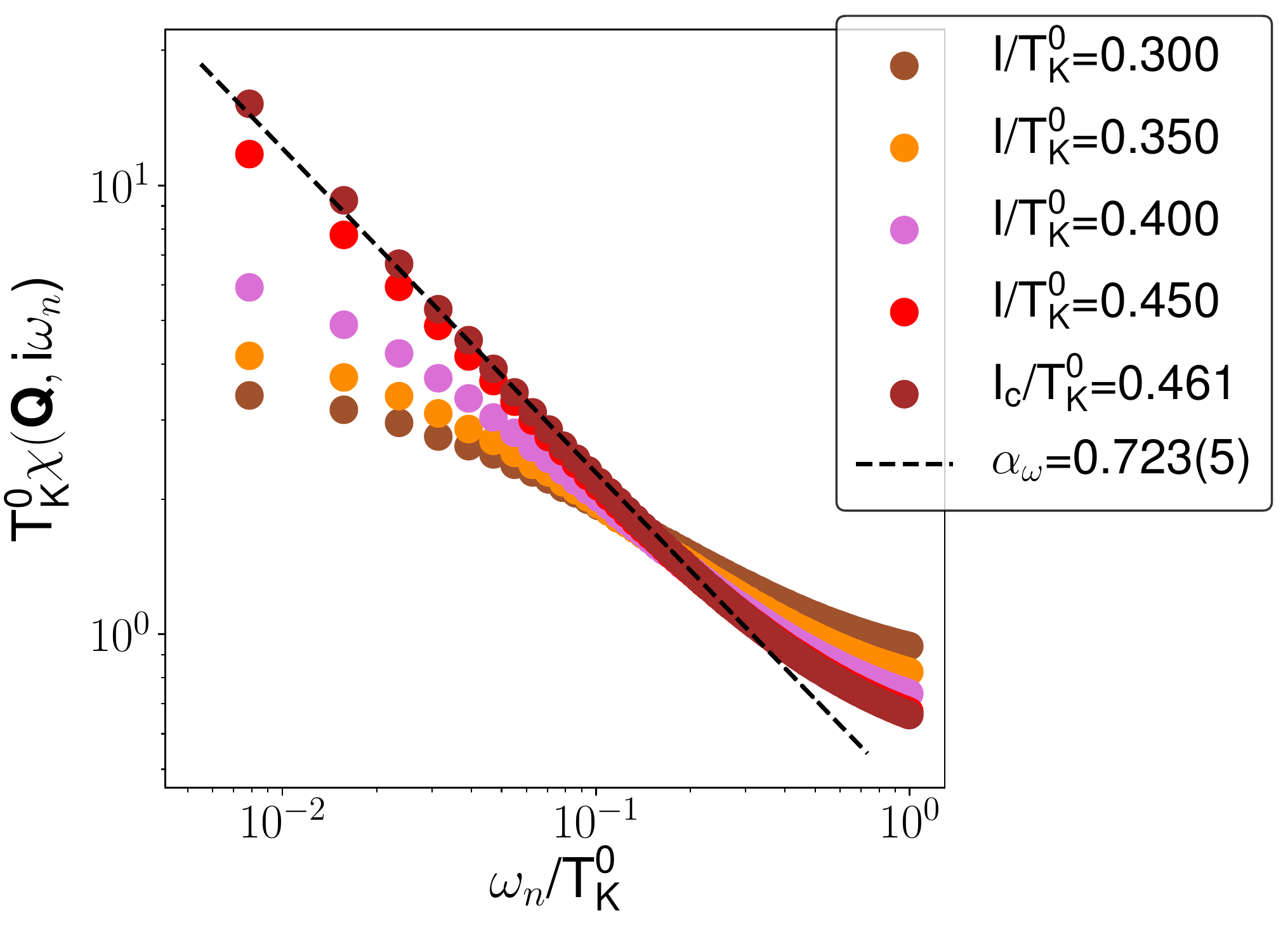}}    
\caption{(Color online) Frequency dependence of the antiferromagnetic lattice spin 
susceptibility at the inverse temperature $ \beta T_K^0 = 800$ }
\label{fig:chi_lat_w}
\end{figure}
In Fig.~\ref{fig:chi_lat_w}, we show the frequency dependence of the antiferromaggnetic lattice spin susceptibility 
upon approaching the QCP from
 the paramagnetic phase.  At the QCP with $I=I_c$, according to the relationship between local and lattice susceptibilities (Eq.~\ref{eq:relation}), the antiferromagnetic lattice susceptibility follows a power-law behavior with 
$
\chi({\bf Q},i\omega_n) \sim {\omega_n^{-\alpha_\omega}} \nonumber 
$. 
Away from the QCP with $I<I_c$, it saturates at low frequencies; this reflects the effect of a nonzero static Kondo singlet amplitude in the 
ground state.

\bibliography{edmft}